\documentclass[useAMS, usenatbib]{mn2e}

\usepackage{url}
\usepackage{float}
\usepackage{graphicx}
\usepackage[english]{babel}
\usepackage{amsmath,amssymb}
\usepackage{fancybox}
\usepackage{booktabs}	

\begin{document}
\def\teff{$T\rm_{eff }$}
\def\kms{$\mathrm {km s}^{-1}$}
\title[Origin of the dark material]{Probing the origin of the dark material on Iapetus}
\author[Tosi et al.]{F. Tosi$^{1}$\thanks{E-mail: federico.tosi@ifsi-roma.inaf.it}\footnotemark[2], D. Turrini$^{1}$\footnotemark[2], A. Coradini$^{1}$, G. Filacchione$^{2}$ and the VIMS Team\\
$^{1}$Institute for Interplanetary Space Physics, INAF, Via Fosso del Cavaliere 100, 00133, Rome, Italy\\
$^{2}$Institute for Space Astrophysics and Cosmic Physics, INAF, Via Fosso del Cavaliere 100, 00133, Rome, Italy}

\date{Accepted XXX. Received XXX; in original form XXX}

\pagerange{\pageref{firstpage}--\pageref{lastpage}} \pubyear{2009}

\maketitle

\label{firstpage}

\begin{abstract}
Among the icy satellites of Saturn, Iapetus shows a striking dichotomy between its leading and trailing hemispheres, the former being significantly darker than the latter. Thanks to the VIMS imaging spectrometer on-board Cassini, it is now possible to investigate the spectral features of the satellites in Saturn system within a wider spectral range and with an enhanced accuracy than with previously available data. In this work, we present an application of the \textit{G-mode} method to the high resolution, visible and near infrared data of Phoebe, Iapetus and Hyperion collected by Cassini/VIMS, in order to search for compositional correlations. We also present the results of a dynamical study on the efficiency of Iapetus in capturing dust grains travelling inward in Saturn system with the aim of evaluating the viability of Poynting-Robertson drag as the physical mechanism transferring the dark material to the satellite. The results of spectroscopic classification are used jointly with the ones of the dynamical study to describe a plausible physical scenario for the origin of Iapetus' dichotomy. Our work shows that mass transfer from the outer Saturnian system is an efficient mechanism, particularly for the range of sizes hypothesised for the particles composing the newly discovered outer ring around Saturn. Both spectral and dynamical data indicate Phoebe as the main source of the dark material. However, due to considerations on the collisional history of the Saturnian irregular satellites and to the differences in the spectral features of Phoebe and the dark material on Iapetus in the visible and ultraviolet range, we suggest a multi-source scenario where now extinct prograde satellites and the disruptive impacts that generated the putative collisional families played a significant role in supplying the original amount of dark material.
\end{abstract}


\footnotetext[2]{These authors equally contributed to this work}

\section{Introduction}
Saturn's third largest satellite, Iapetus, shows one of the most striking dichotomies in the solar system. Observations of the satellite, starting from a couple of years after its discovery by G.D. Cassini in 1671 and culminating in two Voyager flybys and several Earth-based investigations (e.g., \citet{cruik83,squyr83,squyr84}), had shown that the trailing hemisphere has an albedo, composition, and morphology typical of the other icy Saturnian satellites, being highly reflective and spectrally consistent with water ice, while the surface of the leading hemisphere is composed of a very low surface reflectance (geometric albedo 2-6$\%$), reddish material that may include organics and carbon.\\
The origin and nature of the low-albedo hemisphere of Iapetus is still debated, representing one of the most intriguing problems in planetary science. The two main theories of the origin of the low-albedo hemisphere are that it was created by an endogenic geologic process such as flooding by magmas \citep{smith81}, or that it resulted from accretion of exogenous particles \citep{soter74,matt92}. The primary evidence for the endogenic model is the existence of in-filled craters \citep{smith81}, while the main supporting observation for the exogenous model is that the dark material is centred precisely on the apex of motion \citep{matt92}.\\
Disk-integrated spectra collected by the VIMS instrument on-board Cassini during several serendipitous periods in 2004 July and October (at distances between 2 - 3.1 $\times$ 10$^6$ km) enabled a first analysis of Iapetus' surface in the spectral range from 1 to 5 $\mu$m that led to the identification of CO$_2$, which is most prominent on the dark side \citep{bea05a}. These authors also found that a good fit to the low-albedo average spectrum is obtained by a small amount of ice, an even smaller amount of Fe$_2$O$_3$ to account for the ferric absorption band at $\sim$1.0 $\mu$m, 36$\%$ Triton tholin, and substantial amounts of the HCN polymer (``poly-HCN'').\\
On 31 December 2004, the Cassini spacecraft had a distant flyby with Iapetus ($>$ 124,000 km, yielding a maximum spatial resolution on the satellite's surface of $\sim$62 km), that represented the first good chance to acquire spatially resolved spectra of the satellite with better signal-to-noise ratio (SNR) in the range beyond 3 $\mu$m, though in this case the relatively large phase angle (54$^{\circ}$ to 116$^{\circ}$) allowed to explore only a fraction of the dayside, where most of the framed surface was covered by dark material. On these data, \citet{cruik08} reported the identification of a broad absorption band centred at 3.29 $\mu$m, and concentrated in a region comprising about 15$\%$ of the low-albedo surface area, that they interpreted as the C-H stretching mode vibration in polycyclic aromatic hydrocarbon (PAH) molecules. \citet{cruik08} also found, in association with the aromatic band, two weaker bands at 3.42 and 3.52 $\mu$m, attributed to the asymmetric stretching modes of the -CH2- group in aliphatic hydrocarbons. Compounds containing the nitrile (-C$\equiv$N) functional group were initially suspected as the cause of the 2.42-2.44 $\mu$m band, clearly correlated to the low-albedo material on Phoebe, Iapetus and Dione \citep{cruik91,clark05,brown06,clark07}, and possibly some of the indistinct spectral structure in the region 4.5-4.8 $\mu$m \citep{bea05a}; although recently \citet{clark09} revised this interpretation and suggested that molecular hydrogen trapped in dark material would be a better candidate for this feature. From the analysis of far ultraviolet spectra returned by the Cassini \textit{Ultraviolet Imaging Spectrograph} (UVIS), \citet{hah08} found that water ice amounts increase within the dark material away from the apex (at 90 W$^{\circ}$ longitude, the centre of the dark leading hemisphere), consistently with thermal segregation of water ice; yet the fact that water ice is present also at the lowest, darkest and warmest latitudes, where it is not expected to be stable, may be a sign of ongoing or recent emplacement of the dark material from an exogenous source (ibid).\\
Images returned by the \textit{Imaging Science Subsystem} on-board Cassini during this flyby suggested mass wasting of ballistically deposited material \citep{por05}, although the limited spatial resolution did not allow to safely identify any small bright crater lying over the dark terrain, which is a strong argument in favour of deposited material. Moreover, using scatterometry data returned by the \textit{Cassini Titan RADAR Mapper} (named RADAR for concision in the following), \citet{ostro06} pointed out that Iapetus' 2.2-cm (13.78 GHz) radar albedo is dramatically higher on the optically bright trailing side than the optically dark leading side, whereas 12.6-cm (2.38 GHz) results reported by \citet{black04} with the Arecibo Observatory's radar system show hardly any hemispheric asymmetry and give a mean radar reflectivity several times lower than the reflectivity measured at 2.2 cm, which can be explained if the leading side's optically dark contaminant is present to depths of at least one to several decimetres.\\
On September 10, 2007, the Cassini spacecraft performed its first and only targeted Iapetus flyby, at minimum altitude of 1620 km. Approach occurred over the unlit leading hemisphere, departure over the illuminated trailing one. From the ISS data, a strong indication that dark material is lying over bright terrain finally came from the numerous small bright craters detected in the high-resolution images from the flyby. The existence of these small craters (up to some dozens of meters in diameter) indicates that the dark blanket is very thin, probably no more than a few meters, possibly only decimetres. On the other hand, the dark material is expected to be thicker than only millimetres because of the evidence for mass wasting and because of the ability of the Cassini 2.2 cm RADAR to distinguish between the dark and the bright terrain.\\
The confirmation of the exogenous model emphasised the need to identify a reliable source of the dark material. In past times, for different reasons, a number of possible sources were proposed: Phoebe \citep{soter74}, Titan \citep{owen01}, Hyperion \citep{matt92}, or possibly other small dark irregular satellites \citep{bea02,bea05b}, although other works essentially focused on Hyperion and Phoebe as the main candidates. The mechanisms invoked to justify the transfer of material from the source to Iapetus were different for these two bodies: for Hyperion the proposed mechanism was collisional excavation and/or break-up \citep{matt92,mar02,dal04}, while Poynting-Robertson (\emph{PR} in the following) drag was advocated for Phoebe \citep{soter74,bur96}. PR drag was also invoked by \citet{bea05b} in their suggestion that the source of the dark material could be linked to the retrograde irregular satellites as a whole instead than just to Phoebe. Dust ejected from retrograde satellites and migrating inward from the outer Saturnian system would impact Iapetus on its leading hemisphere due to its counter-revolving motion, thus naturally explaining the leading-trailing asymmetry \citep{soter74}. Intuitively, such mechanism should be characterised by a high transfer efficiency since Iapetus' orbital period is shorter than the migration timescale by orders of magnitude, yet to present the only quantitative evaluation of the transfer efficiency is that of \citet{bur96}, who estimated it being about 70$\%$ for 10 $\mu$m sized dust grains. On the contrary, several evaluations of the transfer efficiency from Hyperion to Iapetus have been performed \citep{matt92,mar02,dal04}, the resulting value varying over three orders of magnitude (from 0.1$\%$ to about 20-40$\%$) depending on the characteristics of the ejection process.\\
In addition to the limited information on the transfer efficiency, the major issue of the Phoebe-based model is that the visible spectrum of Iapetus' leading side is similar to D-type asteroids, a primitive, very low albedo group of bodies exhibiting a typically reddish spectrum in the visual region, i.e. increasing in reflectance with increasing wavelength, while Phoebe's spectrum is essentially flat and grey in the visual region, thus similar to C-type or F-type asteroids (see for example \citet{thol83,bea02,grav07}); moreover its albedo is higher than carbon, particularly in the brighter areas detected by Voyager 2 and Cassini.\\
In this paper, we have explored VIMS data returned by the Cassini spacecraft after its close encounter with Iapetus, combining them with VIMS data of Phoebe and Hyperion - acquired with comparable spatial resolution and favourable illumination conditions - that were extensively analysed in previous works. We classify this dataset, with an automatic statistical method, separately for the visual and IR portions, to identify homogeneous types and look for correlations among the spectra of Iapetus, Phoebe and Hyperion. To support the results of the spectral comparison, we reviewed and extended the dust migration scenario based on Poynting Robertson drag in light of the recent results of both theoretical works and the analysis of Galileo, Cassini and Spitzer data. We estimated its mass transfer efficiency by means of a statistical approach derived from the one designed by \cite{kes81} to evaluate the impact probabilities of Jovian irregular satellites. Finally, we discuss the amount of ejecta needed to supply the dark material present on Iapetus and compare it to the one needed in the Hyperion-based scenario and the available observational constrains.

\section{The G-mode method}

The \textit{G-mode} method was originally developed by A.I. Gavrishin and A. Coradini (see \citet{gavr80,gavr92,cora76,cora77}) to classify lunar samples on the basis of the major oxides composition. The good results obtained warranted its application to several different data sets (see, for example, \citet{cora76,carusi78,gavr80,bianchi80,giov81,cora83,barucci87,orosei03}). In particular, the Imaging Spectrometer for Mars (ISM), flown on-board the Soviet Phobos mission, offered the first chance to apply the G-mode method to imaging spectroscopy data \citep{cora91,erard91,cerroni95}. More recently, the G-mode was applied also to Cassini/VIMS data acquired during the close flyby of Phoebe \citep{tosi05,cora08} and to the combined data of Titan returned by Cassini/VIMS and Cassini/RADAR \citep{tosi09}.\\
The G-mode differs from other broadly used statistical methods - such as the \textit{Principal Components Analysis} (PCA) and the \textit{Q-mode} method - in some key characteristics (see \citet{bianchi80}): in summary, a linear dependence of the variables is not needed; instrumental errors can be taken into account; meaningless variables are discerned and removed; and finally different levels of classification can be performed.\\
Basically, by lowering the confidence level of the test, set a priori by the user, the algorithm can perform a more refined classification, in order to look for further homogeneous types. In this case, the G-mode includes a test that allows to interrupt the classification when it becomes too detailed: when the statistical distance among types becomes smaller than the established confidence level, the algorithm can either stop or continue by merging different small types together (this condition is reported in the output of the program).\\
In this specific case, our statistical universe is composed by $N$ spectra sampled in $M$ spectral channels. The G-mode allows the user to apply either the same error value to all the variables (this approach is preferred when all the variables shall drive the classification at the same level), or to assign a specific error value to each variable. In the case of multi-spectral and hyperspectral data, when using reflectances as variables, a logic approach is to use the instrumental noise, that for each spectral channel is given by the inverse of the SNR, as an error. Alternatively, should the depths of some diagnostic absorption bands be used as variables, then the average standard deviation of the band depth can be used as an error for each variable. In this work, we actually apply the first approach to the visible portion of the data, where no absorption band is expected to occur, while the second approach is preferred for the infrared data, where several spectral signatures are used to classify the data.

\section{The VIMS instrument}

The \textit{Visual and Infrared Mapping Spectrometer} (VIMS) is an imaging spectrometer on-board the Cassini Orbiter spacecraft. VIMS is actually made up of two spectrometers, VIMS-V (developed in Italy) and VIMS-IR (developed in the USA). VIMS is the result of an international collaboration involving the space agencies of the United States, Italy, France and Germany as well as other academic and industrial partners.\\
The two channels share a common electronics box and are co-aligned on a common optical pallet. The combined optical system generates 352 two-dimensional images (with maximum nominal dimensions of 64$\times$64, 0.5 mrad pixels), each one corresponding to a specific spectral channel. These images are merged by the main electronics in order to produce ``image cubes'' representing the spectrum of the same field of view (FOV) in the range from 0.35 to 5.1 $\mu$m, sampled in 352 bands. See \citet{brown04,miller96} for a complete description of the instrument.\\
In this work, we separately explored both the visible and infrared portions of the selected cubes.

\begin{table*}
\caption{VIMS specifications summary.}
\centering
\begin{tabular}{lcc}
\toprule
 & \textbf{VIMS-V} & \textbf{VIMS-IR} \\
\midrule
Spectral coverage ($\mu$m) & 0.35 - 1.05 & 0.85 - 5.1 \\
Spectral channels (bands) & 96 & 256 \\
Average spectral sampling (nm/channel) & 7.3 & 16.6 \\
Total FOV ($^{\circ}$) & 1.83$\times$1.83 & 1.83$\times$1.83 \\
Total FOV (mrad) & 32$\times$32 & 32$\times$32 \\
Nominal IFOV (mrad) & 0.50$\times$0.50 & 0.50$\times$0.50 \\
Hi-res IFOV (mrad) & 0.167$\times$0.167 & 0.25$\times$0.50 \\
Nominal image dimension (pixel) & 1$\times$1, 6$\times$6, 12$\times$12, 64$\times$64 & 1$\times$1, 6$\times$6, 12$\times$12, 64$\times$64 \\
Detector type & Si CCD (2D) & InSb photodiodes (1D) \\
Average instrumental SNR & 380 & 100 \\
\bottomrule
\end{tabular}
\label{tab1}
\end{table*}

\section{The dataset}

\begin{figure*}
\resizebox{\hsize}{!}{\includegraphics[clip=true]{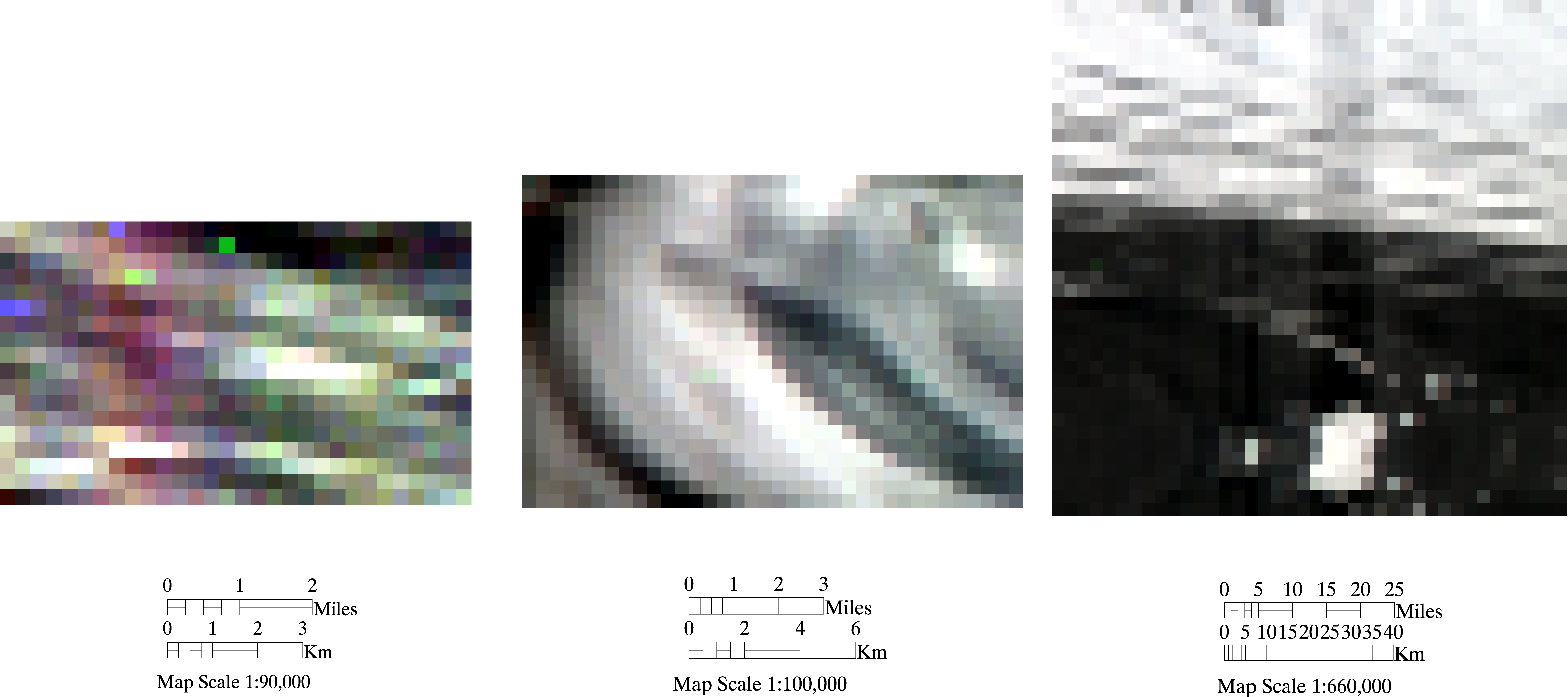}}
\caption{RGB images (R=2.02 $\mu$m, G=1.59 $\mu$m, B=1.28 $\mu$m) of the VIMS cubes used in this work, shown in their IR portion. From left to right, the data refer to Phoebe (cube CM$\_$1465674952), Hyperion (cube CM$\_$1506393701) and Iapetus (cube CM$\_$1568129671), respectively. A horizontal scale bar, placed under each image, accounts for the linear dimension of the image.}
\label{fig1}
\end{figure*}

\begin{figure*}
\resizebox{\hsize}{!}{\includegraphics[clip=true]{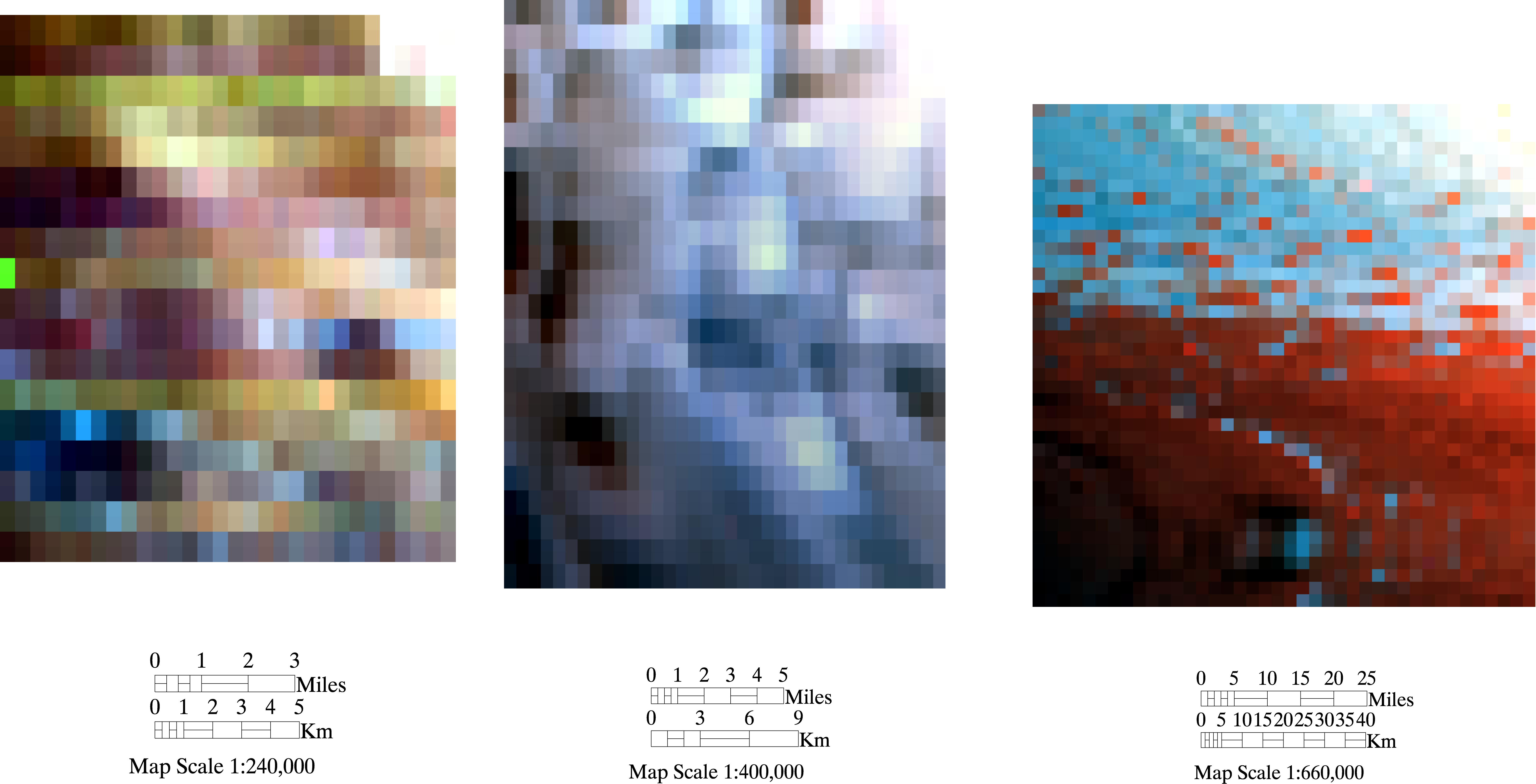}}
\caption{RGB images (R=702.9 nm, G=549.5 nm, B=439.2 nm) of the VIMS cubes used in this work, shown in their VIS portion. From left to right, the data refer to Phoebe (cube CM$\_$1465674952), Hyperion (cube CM$\_$1506393701) and Iapetus (cube CM$\_$1568129671), respectively. A horizontal scale bar, placed under each image, accounts for the linear dimension of the image.}
\label{fig2}
\end{figure*}

Among the VIMS acquisitions, we selected 3 cubes, respectively from the Phoebe flyby and the to-date closest flybys with Iapetus and Hyperion. Such data were selected in order to have: a spatial resolution as high as possible, a low phase angle (intended to select only cubes showing most of targets' dayside, and to limit the dependence on the specific observational geometry), and a good SNR. To do this, we have selected cubes from the three satellites showing at the same time: 1) a spatial resolution $\leq$ 3 km/pixel; 2) a phase angle $\leq$ 43$^{\circ}$; 3) an IR exposure time $\geq$ 120 ms/pixel. Details about these data are given in Tables \ref{tab2} and \ref{tab3}. The VIMS cube selected for Phoebe is the closest available from the flyby, acquired at 19:32 UT on 11 June 2004 at a range of 2096 km and phase angle 24.6$^{\circ}$. Also for Hyperion we selected the closest cube, acquired at 02:15 on 26 September 2005 at a range of 2989 km and a phase angle of 42.6$^{\circ}$. Finally, the cube selected for Iapetus was acquired at 15:01 on 10 September 2007, at a range of 6091 km and a phase angle of 17.3$^{\circ}$: although this cube was not acquired at closest approach but only 40 minutes later, it represents a good tradeoff among image dimensions, spatial resolution, exposure time and phase angle. Moreover, this cube is peculiar because it shows a region located exactly at the boundary between dark and bright terrains, where a relatively sharp division can be seen. In all cases, there are no sky background pixels since the satellite's surface always fills the VIMS FOV. Figure \ref{fig1} shows the RGB images of these cubes in their VIS portion.\\
Due to the fact that the Phoebe and Hyperion cubes were acquired in high resolution IFOV mode both by the visible and infrared channels of VIMS, the IFOV of VIMS-IR being 250$\times$500 $\mu$rad wide while the IFOV of VIMS-V is 167$\times$167 $\mu$rad wide, in the case of these two satellites the IR and VIS images do not match. On the contrary, the Iapetus cube was acquired in nominal IFOV mode (500$\times$500 $\mu$rad) by both channels, so that the IR image basically matches with the VIS image (actually a slight misalignment occurring between the optical axes of the two VIS and IR telescopes prevents a perfect match). Figure \ref{fig2} shows the RGB images of these cubes in their IR portion.
\begin{table*}
\caption{Details of the VIMS cubes used in this work.}
\centering
\begin{tabular}{lclcccccc}
\toprule
\textbf{Satellite} & \textbf{Mission} & \textbf{Dataset} & \textbf{Filename} & \textbf{Date} & \textbf{UTC} & \textbf{Relative} & \textbf{Spacecraft} & \textbf{Phase} \\
 & \textbf{Sequence} & & & & \textbf{on-board} & \textbf{velocity} & \textbf{altitude} & \textbf{angle} \\
 & & & & & & \textbf{(km/s)} & \textbf{(km)} & \textbf{($^{\circ}$)} \\
\midrule
Phoebe & S01 & PHOEBE017 & CM$\_$1465674952 & 11 Jun 2004 & 19:32 & 6.35 & 2096 & 24.576 \\
Hyperion & S14 & HYPERIONC007 & CM$\_$1506393701 & 26 Sep 2005 & 02:15 & 5.64 & 2989 & 42.598 \\
Iapetus & S33 & ORSHIRES001 & CM$\_$1568129671 & 10 Sep 2007 & 15:01 & 2.35 & 6091 & 17.298 \\
\bottomrule
\end{tabular}
\label{tab2}
\end{table*}
\begin{table*}
\caption{Details of the VIMS cubes used in this work.}
\centering
\begin{tabular}{lcccccc}
\toprule
\textbf{Satellite} & \textbf{Image} & \textbf{IR t$_{exp}$} & \textbf{IR Spatial} & \textbf{VIS t$_{exp}$} & \textbf{VIS Spatial} \\
 & \textbf{dimension} & & \textbf{resolution} & & \textbf{resolution} \\
 & \textbf{(px)} & \textbf{(msec/px)} & \textbf{(km/px)} & \textbf{(msec)} & \textbf{(km/px)} \\
\midrule
Phoebe & 30$\times$18 & 180 & 0.524$\times$1.048 & 5630 & 0.349$\times$0.349 \\
Hyperion & 36$\times$24 & 320 & 0.747$\times$1.494 & 7680 & 0.498$\times$0.498 \\
Iapetus & 40$\times$40 & 120 & 3.045$\times$3.045 & 4800 & 1.015$\times$1.015 \\
\bottomrule
\end{tabular}
\label{tab3}
\end{table*}
All of the VIMS cubes used in this work were initially calibrated, in their infrared portion, by means of the RC15 VIMS-IR sensitivity function and the flat-field cube released in 2005, namely the last official products available at the time this work was undertaken. The sensitivity function allows to convert the raw signal of each pixel inside the IR image into radiance (divided by the IR integration time and the flat-field) and then into reflectance $I/F$, where $I$ is the intensity of reflected light (uncorrected for the specific observational geometry and the thermal emission) and $\pi F$ is the plane-parallel flux of sunlight incident on the satellite \citep{theka73}, scaled for its heliocentric distance (for details about the VIMS calibration, see \citet{mccord04}). It should be noted that the latest unofficial RC17 sensitivity function of VIMS-IR, derived in 2008 and currently under test, is different from the RC15 calibration used in \citet{clark05} by the use of additional standard stars and data from the VIMS solar port. It was found that the RC15 calibration, based on comparison to Galileo NIMS data with VIMS data from the Cassini Jupiter encounter, the Moon flyby, plus ground calibration data contained a residual 3-micron absorption and spectral structure in the 2-2.5 micron region due to ringing in the VIMS order sorting filters (Clark, personal communication, 2009). For this reason, in order not to mislead our work, we have separately tested both these sensitivity functions on the VIMS infrared data.\\
Moreover, all the spectra were ``despiked'': we removed single-pixel, single-spectral channel deviations, caused by systematic instrumental artifacts like order-sorting filters as well as random artifacts like cosmic rays or high energy radiation striking the detectors (these events reveal themselves as spikes in the dark current stored in each raw cube). On the visible portion of the VIMS data, also a ``destriping'' procedure was applied, aimed to remove residual offsets between radiometric levels of different CCD columns while observing a uniform scene.\\
For the Phoebe and Hyperion cubes, we selected all the pixels inside the spectral image, so we have $540$ samples from Phoebe and $864$ samples from Hyperion. In the case of the Iapetus cube, showing both dark and bright material, we define a ``region of interest'' including $639$ samples from the dark material, covering $40\%$ of the image. Hence it follows that, by summing the samples from Phoebe, Iapetus and Hyperion, our global dataset is made up by $2043$ samples/spectra.

\section{Discussion}

\subsection{Results in the visible range}

\begin{figure}
\resizebox{\hsize}{!}{\includegraphics[clip=true]{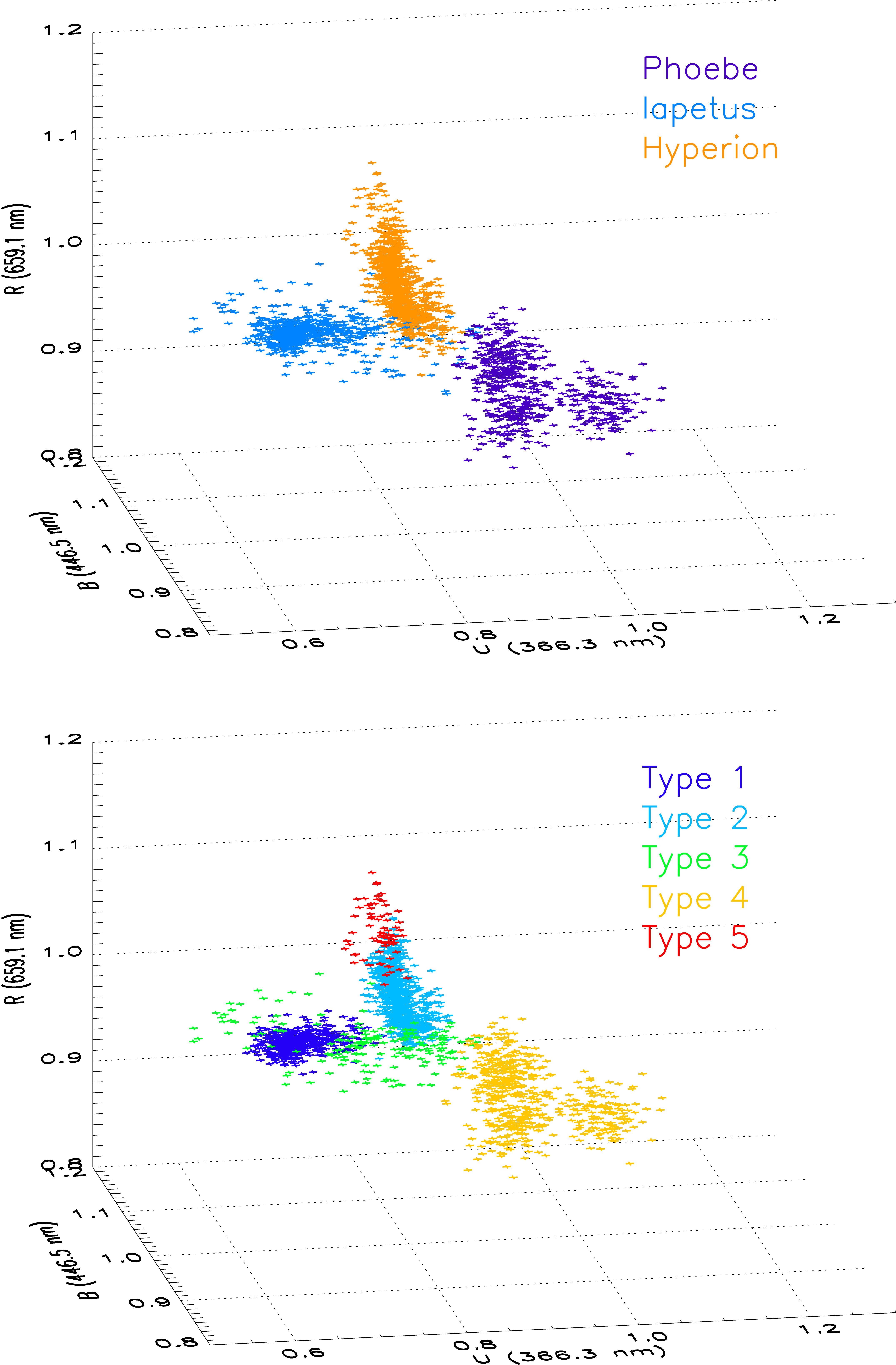}}
\caption{Classification in the visible range. The samples are plotted in a 3D space, with axes corresponding to the normalised reflectances measured in the U, B, and R bands. Upper plot: distribution of the samples as a function of the three satellites. Lower plot: distribution of the homogeneous types returned by the G-mode analysis.}
\label{fig3}
\end{figure}

\begin{figure*}
\includegraphics[height=22cm]{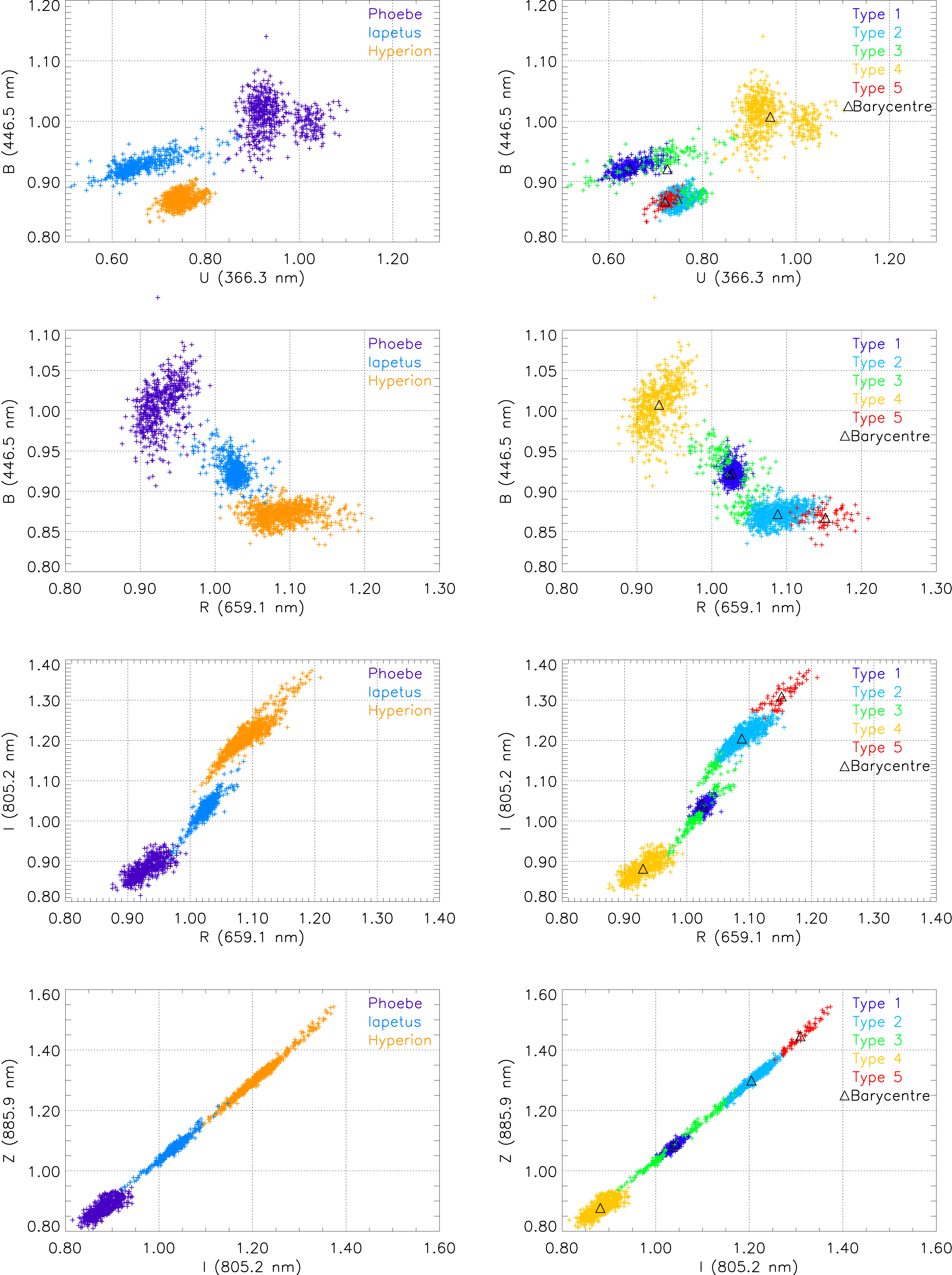}
\caption{Normalised reflectances of the samples from Phoebe, Hyperion and the dark side of Iapetus, projected on the UB, RB, RI, and IZ planes, respectively, from top to bottom. Left column: association with the satellites. Right column: homogeneous types found by the G-mode analysis; a black triangle marks the barycentre of the classes.}
\label{fig4}
\end{figure*}

In the range measured by VIMS-V (0.35 - 1.05 $\mu$m), no signatures can be safely identified in the spectra of Iapetus, Phoebe and Hyperion; hence a spectral classification is driven by the spectral continuum, namely the slope parameter. First, all the 2043 calibrated spectra of Iapetus, Phoebe and Hyperion were normalised to the value of 1.0 at the wavelength of 549.54 nm (VIMS channel 28), then the G-mode was applied with a 97.13$\%$ confidence level (1.90$\sigma$). The classification returned 5 homogeneous types, composed by 505, 744, 175, 538 and 71 samples respectively.\\
A convenient way to represent the results is plotting the data in a multidimensional space with axes corresponding to the VIMS spectral channels nearest to the filters defined in the most common photometric systems: $U$ (366.3 nm), $B$ (446.5 nm), $R$ (659.1 nm), $I$ (805.2 nm) and $Z$ (885.9 nm); the $V$ filter being not considered, as it has been used as a reference wavelength for the normalisation of all the spectra. In this way, the three major types, i.e. type 1, 2 and 4 (in blue, cyan and yellow, respectively), can be easily identified in a 3D space whose axes correspond to the U, B and R photometric variables (see Fig. \ref{fig3}); for better clarity, the same data points can be also plotted in a 2D space with axes corresponding to pairs of the U, B, R, I and Z bands, so that the colour indexes for each sample can be evaluated as well: since we keep the information about the real object (satellite) corresponding to each sample, it can be verified that the main three types correspond to the three satellites (see Fig. \ref{fig4}). Type 1 (blue) is made up by samples of the dark side of Iapetus, that appear rather concentrated in all the photometric variables. Type 2 (cyan) is linked to Hyperion, showing the steeper slope of all three satellites (and thus being redder than the dark side of Iapetus, as inferred by the high reflectance values in the R, I and Z bands) and with samples that show a larger spread with respect to the samples of Iapetus. Type 4 (yellow) is clearly related to Phoebe, whose samples, despite of the small area of the satellite considered for this work, are more scattered in the multidimensional space with respect to the samples of Iapetus and Hyperion (especially towards the shorter wavelengths) and are definitely not reddened but, on the contrary, show subsets of pixels exhibiting a higher reflectance towards the blue and UV wavelengths, consistently with recent Earth-based observations \citep{grav07}. Type 3 (green) includes a relatively large number of samples that show an intermediate behaviour between the dark side of Iapetus and Phoebe or Hyperion. Finally, type 5 (red) is made up by samples of Hyperion that behave differently from the majority of the samples of that satellite, showing higher reflectance values on an average in the R, I and Z bands and thus being redder than all the other types.\\
This classification points out the fact that the colour indexes of the dark side of Iapetus, Phoebe and Hyperion, evaluated on high resolution data, are significantly different from each other. As expected, Hyperion is spectrally red at visible wavelengths, similar to D-type asteroids, with a subset of samples showing the steeper spectral slope of all the dataset; yet this reddening is, on an average, higher than the dark material of Iapetus, while Phoebe represents a totally different case, with no reddening and even a minor number of samples exhibiting a negative slope (i.e. the reflectance increases towards the shorter wavelengths).

\subsection{Results in the near infrared range}

In the range measured by VIMS-IR (0.85 - 5.1 $\mu$m), unlike the visible range case, several absorption features can be identified in the spectra of the three icy satellites. In the past, for classification purposes, we used the spectral range offered by VIMS-IR, both as a whole and truncated at 4.3 $\mu$m in order to discard the contribution of thermal emission for Iapetus, using simple reflectances $I/F$ as variables (see \citet{tosi06}). In this work, we adopted a different approach: we normalise all the spectra to VIMS channel 179, corresponding to the 2.23 $\mu$m wavelength, that is free from absorption features in all the three satellites, then we select a number of diagnostic absorption bands to be used as variables.\\
All the spectra from Iapetus, Phoebe and Hyperion are dominated by the large 3 $\mu$m absorption band due to the OH fundamental stretch in H$_2$O ice and/or bound water, which is therefore not relevant for the classification. Instead, we use the spectral signatures of H$_2$O at 1.51 and 2.05 $\mu$m, combined to a number of spectral features diagnostic for volatiles.\\
Interestingly, the $\nu_3$ asymmetric stretch of CO$_2$ is slightly shifted in position, moving - on an average - from 4.247 $\mu$m on Hyperion to 4.254 $\mu$m on Iapetus and 4.260 $\mu$m on Phoebe \citep{clark05,bea05a,cruik07,fila09}, i.e. a maximum difference of 13 nm, which is anyway comprised into one spectral channel of VIMS-IR (whose spectral resolution at those wavelengths is $\sim$22 nm). We sample this compound at 4.26 $\mu$m, the absorption being broad enough to prevent a significant error to be committed in evaluating the band depth. The aromatic C-H stretch on Phoebe mostly occurs at 3.25 $\mu$m \citep{clark05,cora08} while on Iapetus is reported at 3.29 $\mu$m \citep{cruik08}, i.e. shifted by $\sim$40 nm or two spectral channels of VIMS-IR. We also include the CH aliphatic stretch at 3.41 $\mu$m and its overtone at 1.75 $\mu$m \citep{cruik08}; and the 2.44 broad feature possibly being the overtone of a cyanide compound \citep{clark05} or the fundamental of molecular hydrogen trapped in dark material \citep{clark09}. According to the first interpretation, the fundamental of CN varies in position as a result of variations in its chemical environment; assuming the RC15 calibration, in the Phoebe spectrum a feature is seen in the 4.5-4.6 $\mu$m range \citep{clark05} and we evaluate this bond at 4.53 $\mu$m. Finally, assuming the RC15 calibration we have also considered the 4.42 $\mu$m signature, reported by \citet{cora08} in homogeneous types identified on the surface of Phoebe and suggeted to be related to a nitrile compound like HC$_3$N or HNCO. It should be noted that by applying the RC17 calibration, the 4.53 $\mu$m and 4.42 $\mu$m features tend to disappear: such a behaviour can be indicative of calibration artifacts, so we discarded these two variables in our second test.\\
The list of the spectral features used as variables in our classification is summarised in Table \ref{tab4}. For each spectral signature, we compute the band depth following the definition by \citet{clark84}, as:

\begin{equation}
	D = 1 - \frac{R_b}{R_c}
\end{equation}

where \textit{R$_b$} is the reflectance measured at the band centre and \textit{R$_c$} is the reflectance of the spectral continuum at the band centre, reconstructed through a linear fit relying on the wings of the band.\\
We also combine these band depths to the normalised reflectance of the spectral continuum sampled in two wavelengths that are safely free from absorptions (at 1.82 $\mu$m and 3.55 $\mu$m), as a set of convenient variables to be used for the classification (see Table \ref{tab4}).

\begin{table*}
\caption{List of spectral features used to classify the VIMS data in the infrared range.}
\centering
\begin{tabular}{cccl}
\toprule
\textbf{Variable name} & \textbf{VIMS Spectral channel} & \textbf{Wavelength ($\mu$m)} & \textbf{Compound/Spectrally active bond} \\
\midrule
BD1 & 135 & 1.51 & H$_2$O ice \\
BD2 & 150 & 1.75 & C-H stretch overtone \\
BD3 & 168 & 2.05 & H$_2$O ice \\
BD4 & 191 & 2.44 & C$\equiv$N overtone or trapped H$_2$ \\
BD5 & 240 & 3.25 & aromatic C-H stretch \\
BD6 & 250 & 3.41 & aliphatic C-H stretch \\
BD7 & 301 & 4.26 & CO$_2$ \\
BD8 & 310 & 4.42 & Nitrile? (RC15 calibration only) \\
BD9 & 317 & 4.53 & C$\equiv$N fundamental stretch (RC15 calibration only) \\
CL1 & 154 & 1.82 & Continuum \\
CL2 & 258 & 3.55 & Continuum \\
\bottomrule
\end{tabular}
\label{tab4}
\end{table*}

\begin{figure}
\resizebox{\hsize}{!}{\includegraphics[clip=true]{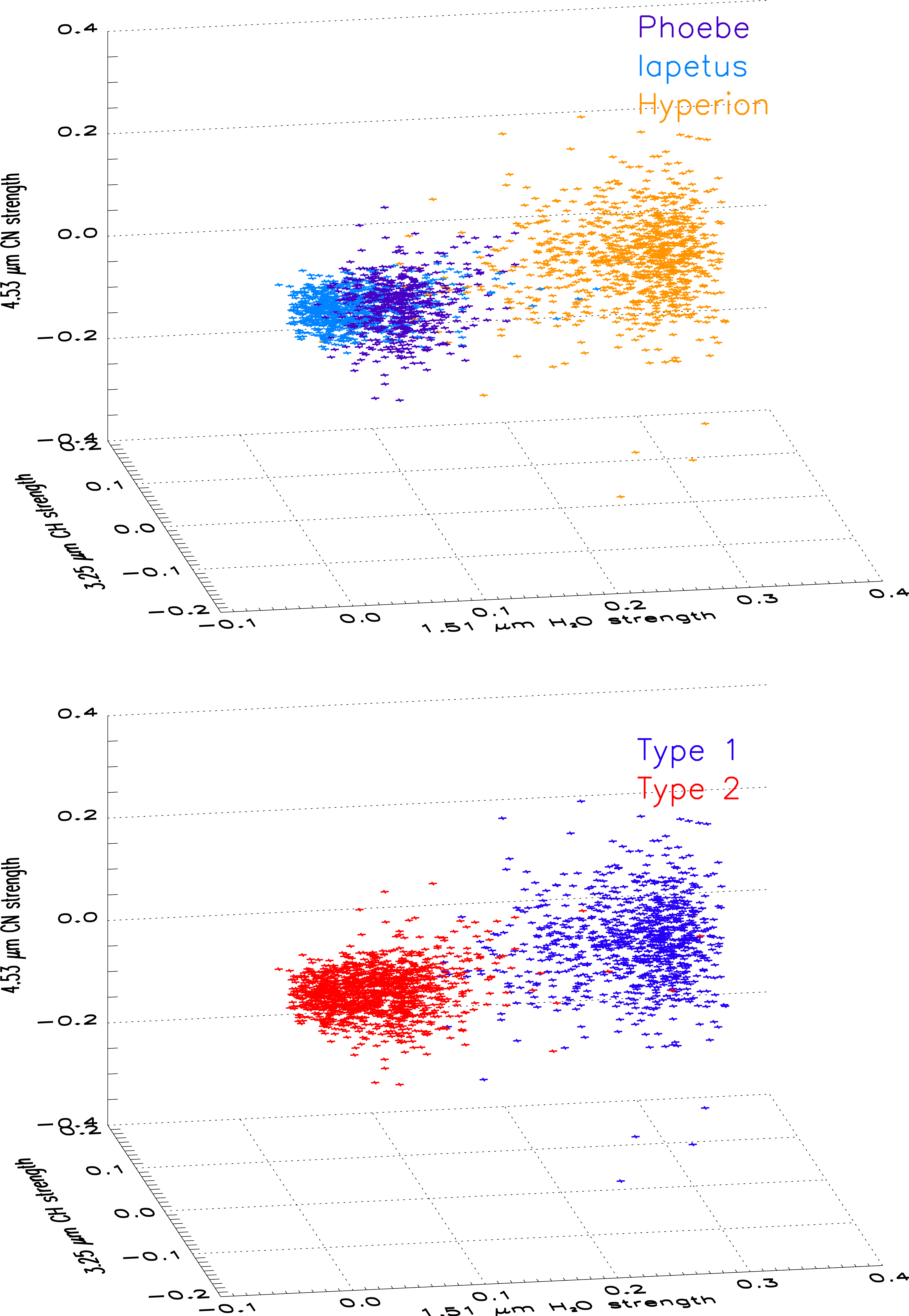}}
\caption{Classification in the near infrared range. The samples are plotted in a 3D space, with axes corresponding to the band depths of water ice (1.51 $\mu$m), CO$_2$ (4.26 $\mu$m) and the CN fundamental stretch (4.53 $\mu$m). Upper plot: distribution of the samples as a function of the three satellites. Lower plot: distribution of the homogeneous types returned by the G-mode analysis.}
\label{fig5}
\end{figure}

By composing such a dataset and classifying it through the G-mode method with a 98.98$\%$ confidence level (2.05$\sigma$), we find 2 types made up of 849 and 1194 samples, respectively. By plotting the result in a 3D space with axes corresponding, for example, to the abundances of H$_2$O ice (measured at 1.51 $\mu$m), CO$_2$ (measured at 4.26 $\mu$m) and C$\equiv$N (measured at 4.53 $\mu$m), these types can be highlighted and the corresponding samples can be put in association with the satellite (see Fig. \ref{fig5}).

\begin{figure*}
\includegraphics[height=22cm]{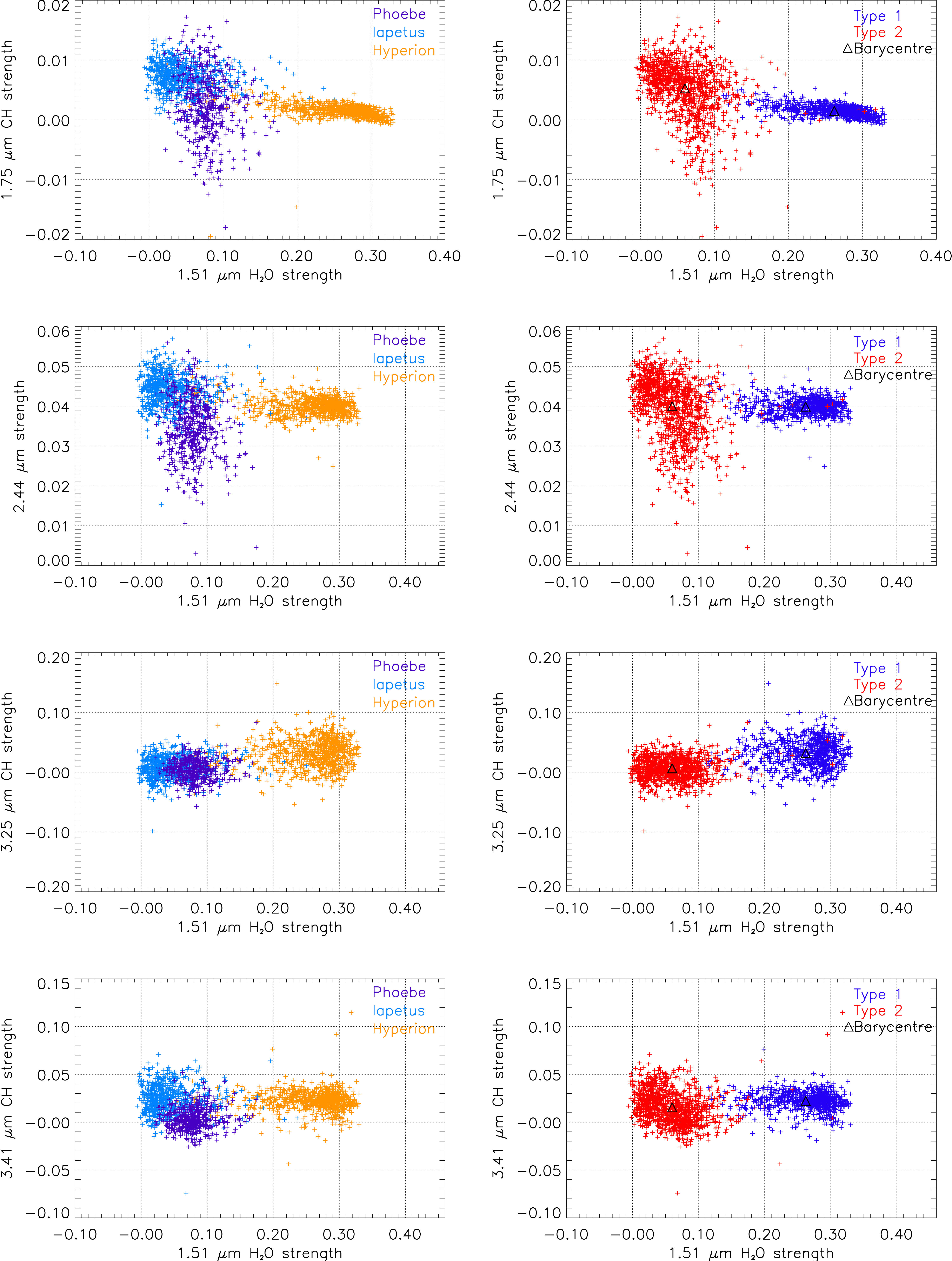}
\caption{Samples from Phoebe, Hyperion and Iapetus, calibrated through the RC15 (2005) sensitivity function and projected on the different planes representing, respectively from top to bottom, the band depths of: the CH overtone signature at 1.75 $\mu$m, the 2.44 $\mu$m feature, the aromatic CH stretch at 3.25 $\mu$m, and the CH stretch at 3.41 $\mu$m, with respect to the 1.51 $\mu$m water ice band depth. Left column: association with the satellites. Right column: homogeneous types found by the G-mode analysis; a black triangle marks the barycentre of the classes.}
\label{fig6}
\end{figure*}

\begin{figure*}
\includegraphics[height=22cm]{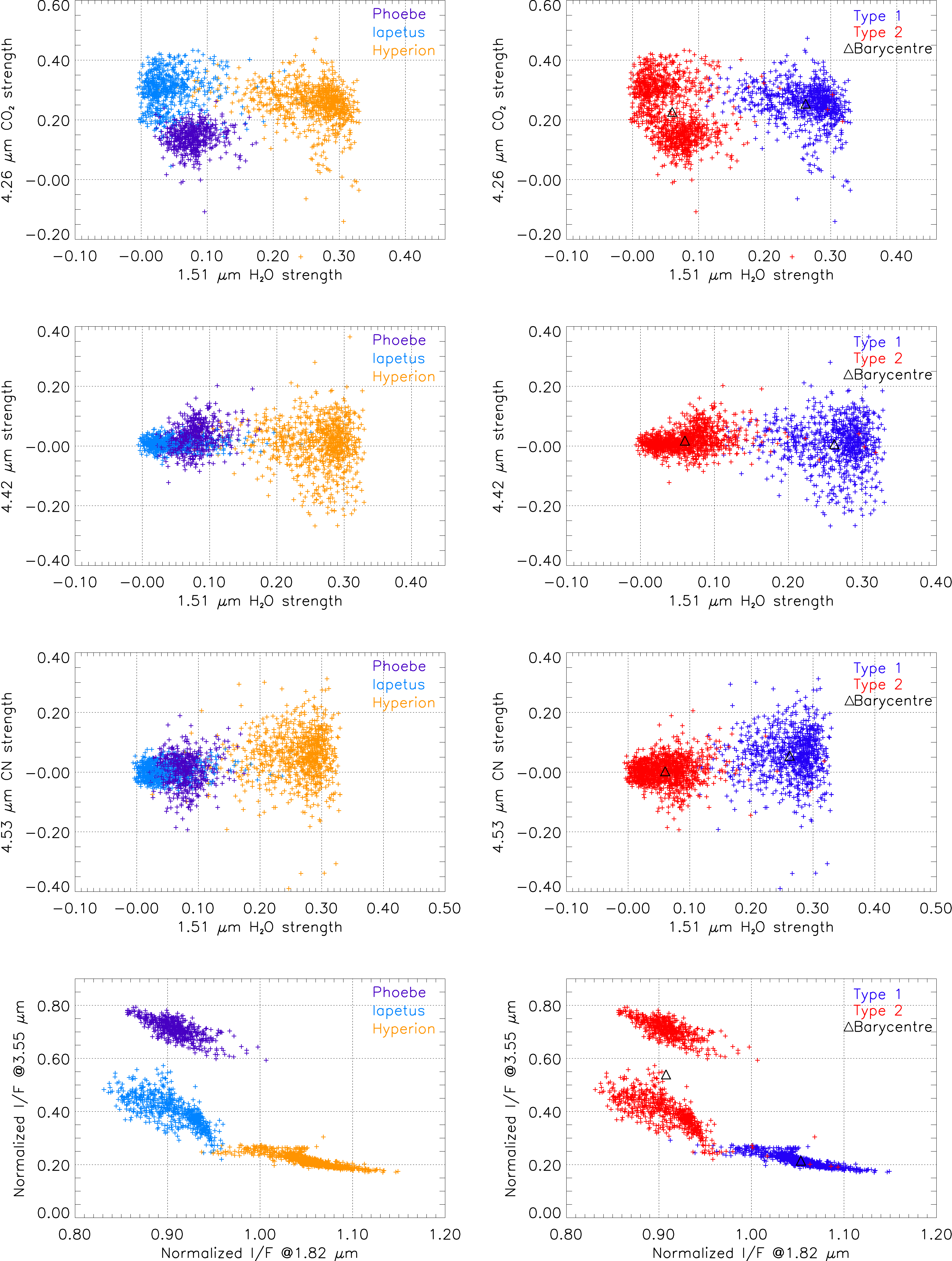}
\caption{Samples from Phoebe, Hyperion and Iapetus, calibrated through the RC15 (2005) sensitivity function and projected on the different planes representing, respectively from top to bottom, the band depths of: CO2 at 4.26 $\mu$m, the CN-related feature at 4.42 $\mu$m, the CN fundamental stretch at 4.53 $\mu$m, and the spectral continuum profile (normalised I/F at 3.55 $\mu$m versus normalised I/F at 1.82 $\mu$m). Left column: association with the satellites. Right column: homogeneous types found by the G-mode analysis; a black triangle marks the barycentre of the classes.}
\label{fig7}
\end{figure*}

For clarity, the same points can be also plotted in a 2D space with axes corresponding to the depths of some diagnostic spectral features (aromatic/aliphatic CH, CO$_2$, C$\equiv$N, etc.) versus the 1.51 $\mu$m signature, related to the abundance of water ice. Keeping the information about the satellite connected to each sample, it can be easily verified that samples of type 1 (blue) entirely correspond to spectra of Hyperion, while samples of type 2 (red) are - with the exception of only one sample - spectra of Phoebe and of the dark side of Iapetus (see Figs. \ref{fig6} and \ref{fig7}).\\
From the classification, it turns out that Hyperion and Phoebe are generally richer in water ice than the Iapetus dark terrain, consistently with the results obtained in the far ultraviolet wavelengths by Cassini/UVIS \citep{hah08}, with a subset of the samples of Iapetus and Phoebe exhibiting similar amounts of ice. The abundance of CO$_2$ is more prominent in the Iapetus dark material than on Phoebe, in agreement with the results by \citet{bea05a}, with Hyperion representing an intermediate situation and showing a stronger correlation with water ice. The abundances of hydrocarbons (traced by the CH aromatic and aliphatic stretches) are quite similar on the three bodies, but with a stronger correlation with water ice in the case of Hyperion. On the other hand, the dark material of Iapetus and Phoebe show similar amounts of nitrile compounds (assuming CN-bearing molecules as traced by the 2.44, 4.42 and 4.53 $\mu$m signatures), that however turn out again to be deeply bounded to water ice on Hyperion, whereas such a correlation is not found on the other two bodies.\\
Regarding the spatial variability, for the considered dataset it should be noted that CO$_2$ is more scattered in samples from Iapetus and Hyperion than in samples of Phoebe. Nitriles traced at 4.53 $\mu$m and 4.42 $\mu$m show a larger variability on Hyperion, although the opposite is interestingly observed for the 2.44 $\mu$m feature, especially on Phoebe. The aromatic CH stretch is more variable on Hyperion, whereas the aliphatic CH signature shows a larger variability on Iapetus. Finally, the spectral continuum sampled at 1.82 $\mu$m (once the spectra have been normalised at 2.23 $\mu$m) turns out to be similar for Phoebe and Iapetus, while Hyperion shows a higher I/F at that wavelength; the three satellites show anyway different I/Fs at 3.55 $\mu$m.

\begin{figure*}
\includegraphics[height=17.4cm]{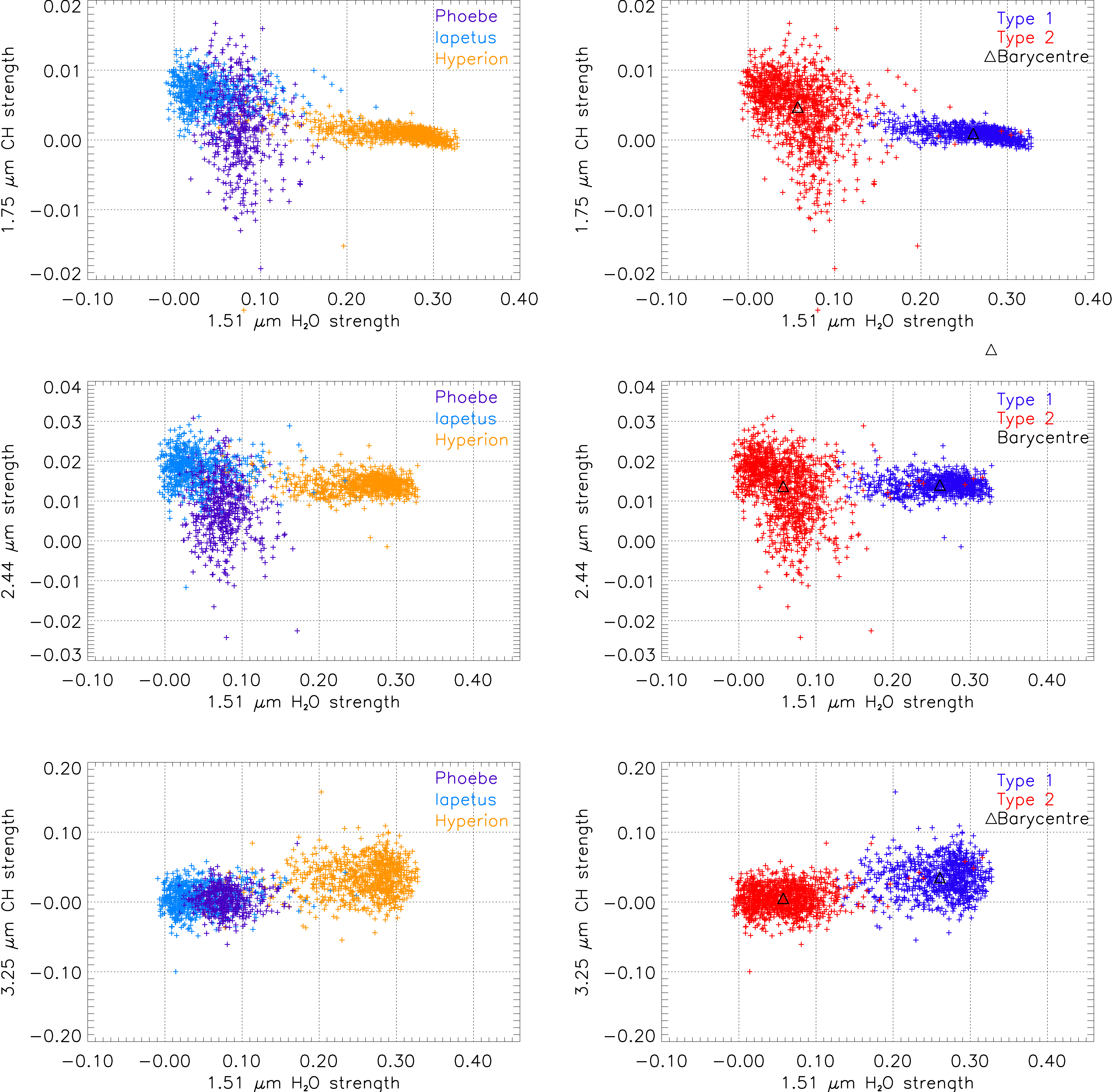}
\caption{Samples from Phoebe, Hyperion and Iapetus, calibrated through the RC17 (2008) sensitivity function and projected on different planes representing, respectively from top to bottom, the band depths of: the CH overtone signature at 1.75 $\mu$m, the 2.44 $\mu$m feature, and the aromatic CH stretch at 3.25 $\mu$m with respect to the 1.51 $\mu$m water ice band depth. Left column: association with the satellites. Right column: homogeneous types found by the G-mode analysis; a black triangle marks the barycentre of the classes.}
\label{fig8}
\end{figure*}

\begin{figure*}
\includegraphics[height=17.4cm]{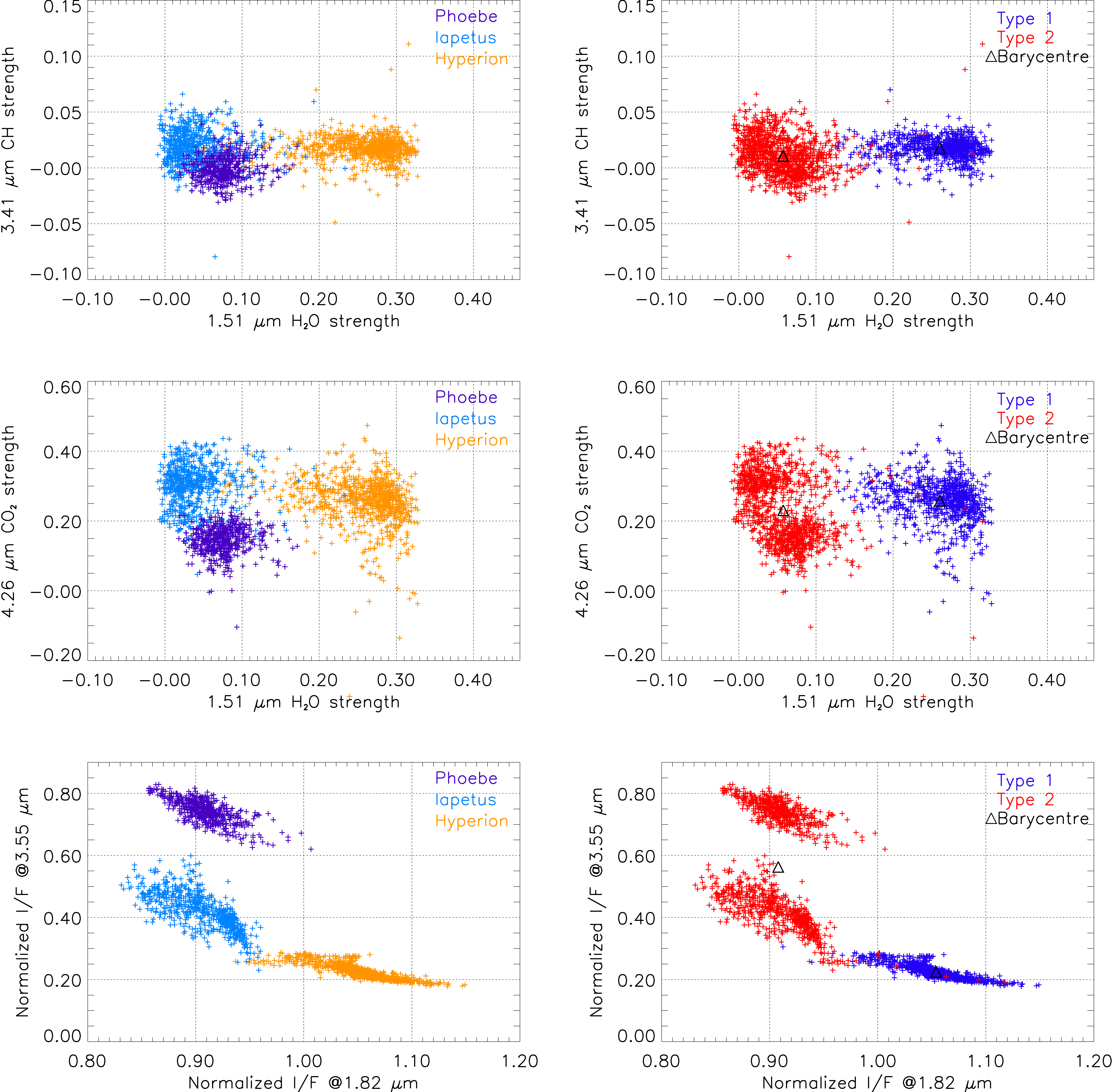}
\caption{Samples from Phoebe, Hyperion and Iapetus, calibrated through the RC17 (2008) sensitivity function and projected on different planes representing, respectively from top to bottom, the band depths of: the CH stretch at 3.41 $\mu$m, CO2 at 4.26 $\mu$m, and the spectral continuum profile (normalised I/F at 3.55 $\mu$m versus normalised I/F at 1.82 $\mu$m). Left column: association with the satellites. Right column: homogeneous types found by the G-mode analysis; a black triangle marks the barycentre of the classes.}
\label{fig9}
\end{figure*}

After this analysis, we redid the work in the infrared range by applying the latest, unofficial sensitivity function derived for VIMS-IR (RC17, Roger Clark, personal communication, 2008). As mentioned before, with this new calibration the features at 4.42 and $\sim$4.5 $\mu$m disappear on most data, so they are no longer considered as variables for the classification, while we keep all the other features, so that the new dataset is made up by 2043 spectra in 9 variables. By classifying it through the G-mode method with the same confidence level used before (98.98$\%$ or 2.05$\sigma$), we find 2 types made up of 844 and 1199 samples, respectively. Figures \ref{fig8} and \ref{fig9} show the result of the classification, with samples represented in a 2D space and axes again corresponding to the depths of the considered spectral features with respect to the 1.51 $\mu$m signature related to the abundance of water ice. It can be seen that the classification remains essentially unchanged with respect to the application of the RC15 sensitivity function, with the same variability of the key compounds on the satellites and a general strong correlation with water ice that still characterizes the samples of Hyperion in comparison to those of Iapetus and Phoebe.\\
In light of this analysis, a general evidence is that, regardless of the specific sensitivity function used to calibrated the VIMS data, the situation in the near infrared range is different with respect to the visible range: on the basis of the depths of either nine or seven spectral features, related to the abundances of several compounds or spectrally active bonds on the surfaces of the satellites, and of the spectral continuum profile sampled in two different wavelengths, the existence of two classes defines a general higher degree of similarity between the dark hemisphere of Iapetus and Phoebe; the G-mode method allows to come to this conclusion through an automatic statistical approach applied to a large number of samples, thus increasing the degree of confidence. However, despite the small spatial scale of the considered data, this analysis points out a general stronger correlation with water ice for Hyperion, and a certain degree of variability of spectral features related to hydrocarbons, CO$_2$ and other compounds on the three satellites.

\section{Origin of the dust: physical and dynamical scenario}

As we mentioned in the introduction, Phoebe and Hyperion have historically been proposed as the main potential sources of the dark material coating the leading hemisphere of Iapetus \citep{soter74,matt92}. More recently, \citet{bea05b} suggested that the source of this material could be linked to more than one object, namely the retrograde Saturnian irregular satellites. To delivery the dark material from Hyperion to Iapetus-crossing orbits, dynamical models relied on high ejection velocities (generally $v$ $>$ 100 m s$^{-1}$, see \citet{matt92,mar02,dal04}) and possibly on collimated clouds of ejecta \citep{matt92,mar02}. PR drag, causing the dust grains to spiral inward toward Saturn, was instead invoked by those scenarios linking the dark material to Phoebe and the irregular satellites \citep{soter74,bea05b}.\\
In this section we will discuss the dust production and transfer mechanisms in the outer Saturnian system (i.e. the region populated by the irregular satellites) and compare them with the ones estimated for the Hyperion-based scenario.

\subsection{Irregular satellites: impacts and dust production}

The images of Phoebe taken by Cassini mission \citep{por05} supply the first evidence that impacts played an important role in the history of the irregular satellites of Saturn. Prior to the arrival of Cassini, \citet{nes03} predicted that collisional removal between the population of irregular satellites should have caused significant modifications of the Saturnian system over the age of the Solar System. Phoebe, as the biggest member of the Saturnian irregular satellites, likely played a major role in this removal process (ibid). At the time, only 13 irregular satellites had been discovered around Saturn: estimating the collisional evolution of this population, \citet{nes03} reported that about 6 impacts (mostly due to Phoebe) were to be expected over a time comparable to the age of the Solar System. Using an updated sample counting 35 irregular satellites, \citet{tur08} showed that, taking into account the present population, a number of impacts more than twice higher should be expected over a time of 3.5$\times$10$^{9}$ years, i.e. the hypothesised lower limit to the age of the irregular satellites. Such results would imply an original population at least 33$\%$ higher than the present one (ibid). Moreover, the orbital structure and the radial distribution of the Saturnian irregular satellites suggests that Phoebe played a major role in clearing its near-by orbital region from once existing collisional shards and smaller objects \citep{tur08,tur09}. Finally, even if less clearly than the Jovian system, the orbital structure of the Saturnian irregular satellites hints to the existence of possible collisional families produced by the disruption of bigger parent bodies \citep{nes03,tur08}.\\
Most impacts between Saturnian irregular satellites would involve pairs of counter-revolving, i.e. prograde vs. retrograde, objects \citep{nes03,tur08}. As shown by \citet{nes03}, the associated impact velocities are about $3$ km s$^{-1}$. This implies that the dust-generating collisions in the Saturnian system would be as energetic as those in the Jovian system, which mainly involve members of the sole prograde or retrograde populations of irregular satellites and have impact velocities of the order of $1-2$ km s$^{-1}$ (ibid). Also collisions between the Saturnian irregular satellites and interplanetary objects, likely at the basis of the formation of the putative collisional families, would be characterised by the same range of impact velocities. As shown by \citet{zea03}, collisions of ecliptic comets on Phoebe would involve average impact velocities of about $3$ km s$^{-1}$. Since the impact velocity is mainly due to the encounter velocity between the host planet and the comet (ibid) and is little influenced by the differences in the small orbital velocities of the irregular satellites, we can assume the value computed by \citet{zea03} for Phoebe as generally valid for all Saturnian irregular satellites.\\
In addition to these possible major impact events, it's been suggested by \citet{kri02} that impacts of interplanetary micrometeorites with small, atmosphereless bodies like the irregular satellites would supply a smaller yet continuous dust source. Such hypothesis is based on the data collected by Galileo space mission, which suggest that a dust production process is still ongoing in the system and is responsible for an enhancement of the circum-Jovian dust respect to interplanetary dust of about an order of magnitude (ibid). While the results of \citet{kri02} refer exclusively to Jovian system, their implications are in principle valid for all giant planets with the caveat that smaller fluxes of micrometeorites are to be considered. As shown by \citet{zea03} for ecliptic comets, in fact, the impact rate on Phoebe is on average half than that on Himalia, the biggest Jovian irregular satellite. As showed by both theoretical modeling (see e.g. \citet{caf97,ben99}) and observations (e.g. \citet{mea05,iaa08} and references therein for ejecta speed from Deep Impact experiment), most of the produced dust particles would be ejected with velocities lower than those (i.e. $v$ $\gtrsim$ 400 m s$^{-1}$) needed to escape Saturn's gravitational attraction \citep{tur09}, thus remaining on planetocentric orbits.\\
While all these dust production processes were suggested by indirect evidences and comparative considerations with the Jovian system, the recent discovery of a ring of particles around Saturn spanning over most of the orbital region of the irregular satellites \citep{vsh09} represents the proof that dust production processes acted and, by analogy with the Jupiter's gossamer rings, are still acting in the Saturnian system (ibid). Even if the information on this newly discovered disk is still limited, we argue that its formation is likely connected to the capture of Phoebe \citep{tur09} and the origin of Phoebe's gap \citep{tur08,tur09}. If collisional families are present between the Saturnian irregular satellites as suggested by \citet{nes03} and \citet{tur08}, the disruptive collisions that generated them would supply additional dusty material and collisional shards which are longer-lived than those ejected in the orbital region crossed by Phoebe \citep{tur08}. 


\subsection{Transfer mechanism from the outer Jovian system}

Dust grains in orbit around Saturn would experience various perturbing effects, namely the gravitational perturbations of the Sun and the outer planets, the radiation pressure and, on longer timescales, the PR drag. While particles smaller than $1.5$ $\mu$m would be rapidly ejected from the Saturnian system due to the effects of radiation pressure as reported by \citet{vsh09}, bigger particles (i.e. $>3.5$ $\mu$m, ibid) would likely survive long enough for PR drag to act. Following the work by \citet{soter74} and \citet{bur96}, the orbital evolution of such dust particles would cause them to spiral towards Saturn. If the dust grains where originally located outside the orbit of Iapetus, during their radial migration they intersect the orbit of the satellite. During their inward drift, dust grains moving on retrograde orbits would impact Iapetus on the leading hemisphere due to their counter-revolving motion, while the same argument does not apply to prograde grains. Moreover, as a consequence of the counter-revolving motion, retrograde dust particles would experience a higher frequency of close encounters with the satellite, suggesting that Iapetus should be more efficient in collecting them than those in prograde motion. This discrepancy in the capture efficiency could justify the asymmetry in the distribution of the dark material covering the surface of the satellite. Such mass transport mechanism is probably still active in the Solar System, as shown by the results of Galileo mission we mentioned above. The data supplied by Galileo spacecraft, moreover, indicated that a significant fraction of the collected dust grains were on retrograde planetocentric orbits \citep{kri02}.\\
To evaluate the efficiency of Iapetus to collect the dust grains drifting towards Saturn due to Poynting-Robertson drag, we used a modified version of the algorithm developed by \citet{kes81}. The original algorithm has been designed to study the collisional evolution of the Jovian irregular satellites by computing the impact probabilities between pairs of them. The algorithm takes into account the spatial and temporal density distributions of the two satellites, assuming that their main orbital elements (semimajor axis, eccentricity and inclination) are fixed in time and that in the timespan considered their remaining orbital angles would sample uniformly all possible values (i.e. the timespan considered should be longer than their precession timescales). While the dust grains migrate inward, they go subject to the gravitational perturbations of Jupiter and the Sun in the outer Saturnian system and of the regular satellites in the inner Saturnian system. As a consequence, the invariance of the main orbital elements cannot be assumed.\\
To overcome this issue, instead of evaluating the impact probability between Iapetus and real, migrating and evolving dust grains, we used Kessler's method to evaluate the efficiency of Iapetus in sweeping different regions of the orbital phase space. We considered a radial region comprised in the range [1.25$\times$10$^{-2}$ - 2.38$\times$10$^{-1}$] AU from Saturn (i.e. the radial region where orbits with different eccentricity values can intersect the one of Iapetus) divided into 20 concentric rings. In this region we computed the probability of collision of Iapetus with synthetic massless particles whose eccentricity and inclination\footnote{The inclination of the particle's orbit is measured respect to the planet's orbital plane about the Sun} values vary in the ranges [0-0.9] and [0$^{\circ}$-60$^{\circ}]$ for prograde orbits or [120$^{\circ}$-180$^{\circ}$] for retrograde orbits. The sampling steps assumed were respectively 0.1 and 2$^{\circ}$. The time over which to integrate the impact probability for each particle has been assumed equal to the time needed to cross the radial ring where it is initially located. This migration time is computed through the relationship for planetocentric Poynting-Robertson drag (see p. 430 from \citet{dpl01}):
\begin{equation}
 \frac{da}{dt}=-\frac{a}{t_{pr}}\frac{5+cos^{2}i_{*}}{6}
\end{equation}
where $t_{pr}$ is a characteristic decay time given by 
\begin{equation}
 t_{pr}=\frac{1}{3\beta}\frac{r^{2}_{\odot}}{GM/c}\approx530\frac{r^{2}_{AU}}{\beta}\,yr,
\end{equation}
$\beta$ is the ratio between forces due to the radiation pressure and the gravity of the Sun, $a$ is the semimajor axis of the particle, $i$ its inclination respect to the orbital plane of the planet about the Sun, $r_{\odot}$ is the Sun-Saturn distance and $r_{AU}$ is the same distance expressed in astronomical units. Inclination and eccentricity are not influenced by PR drag (ibid). As our template orbit for Iapetus we used the mean orbital elements supplied by the JPL Solar System Dynamics website\footnote{\url{http://ssd.jpl.nasa.gov/?sat_elem}}. However, since the inclination referred to the local Laplace plane while our reference frame is the orbital plane of Saturn, we recomputed the mean inclination of the satellite ($i$=10.152$^{\circ}$) from the results of Model 2 of \citet{tur08}.\\
For our numerical setup and the grain sizes we considered (see section \ref{efficiency}), the migration times are always longer (usually by at least an order of magnitude) than the precession timescales of the bodies involved: this allowed us to apply Kessler's method to each radial region. To estimate the impact probability of real dust grains, we computed the cumulative impact probability over the whole radial path of particles sharing the same eccentricity and inclination values. Globally, we considered 6510 prograde orbits and 6510 retrograde ones: once integrated over the radial path, the number of orbits accounted for is 310 for each of the two cases in inclination.\\

\subsection{Dust capture efficiency}\label{efficiency}

\begin{figure}
\centering
\resizebox{\hsize}{!}{\includegraphics[clip=true]{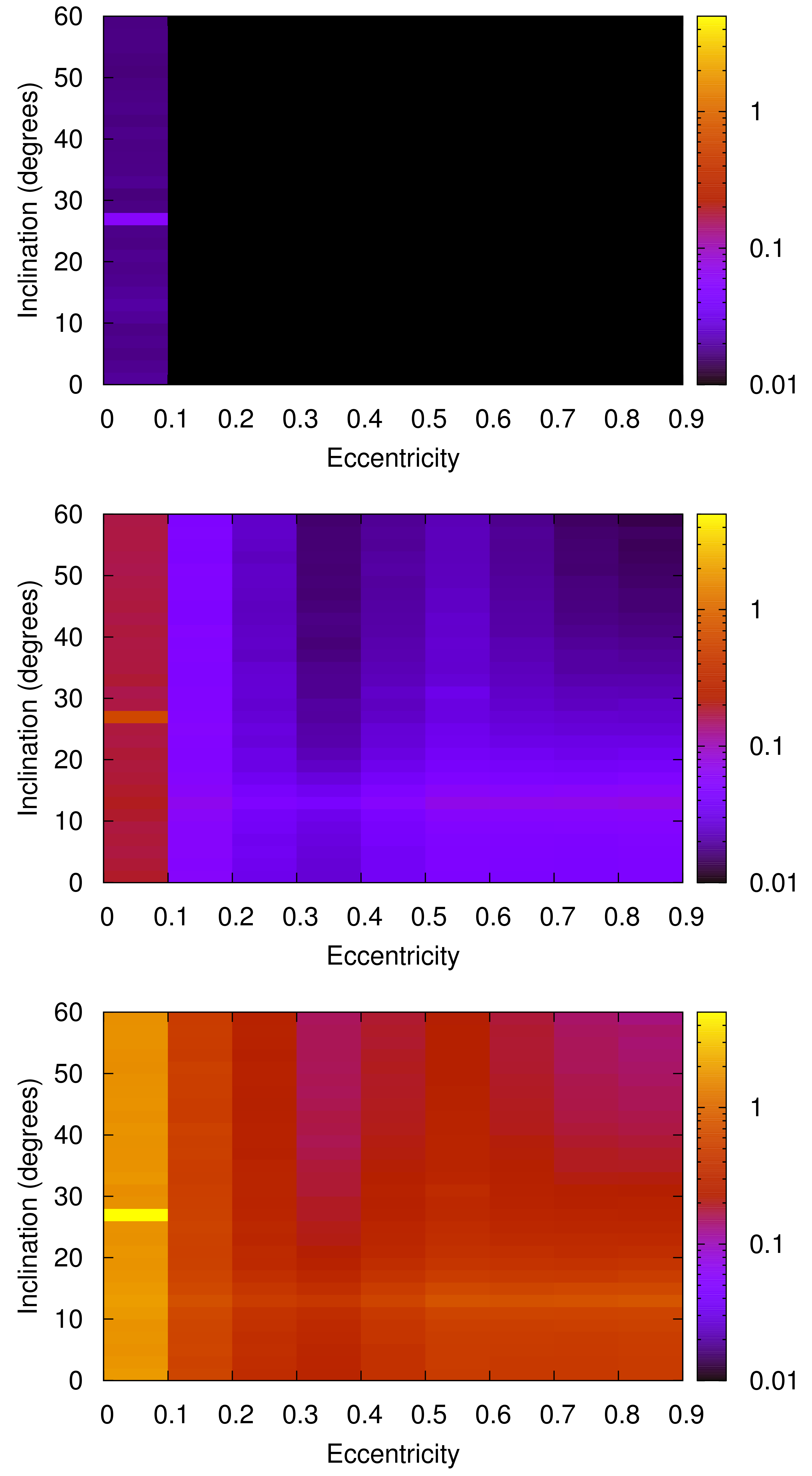}}
\caption{Dust capture efficiency of Iapetus for increasing sizes of the grains (from top to bottom: 0.1, 1 and 10 microns) orbiting Saturn on prograde orbits. Capture efficiency is expressed as the number of impacts over the migration time across the whole radial path. Dust grains ejected by prograde satellites would likely reside in the region of the $e-i$ plane ($30^{\circ} < i < 60^{\circ}$ and $e > 0.1$) where Iapetus' capture efficiency is lowest.}
\label{cloud-pro}
\end{figure}

\begin{figure}
\centering
\resizebox{\hsize}{!}{\includegraphics[clip=true]{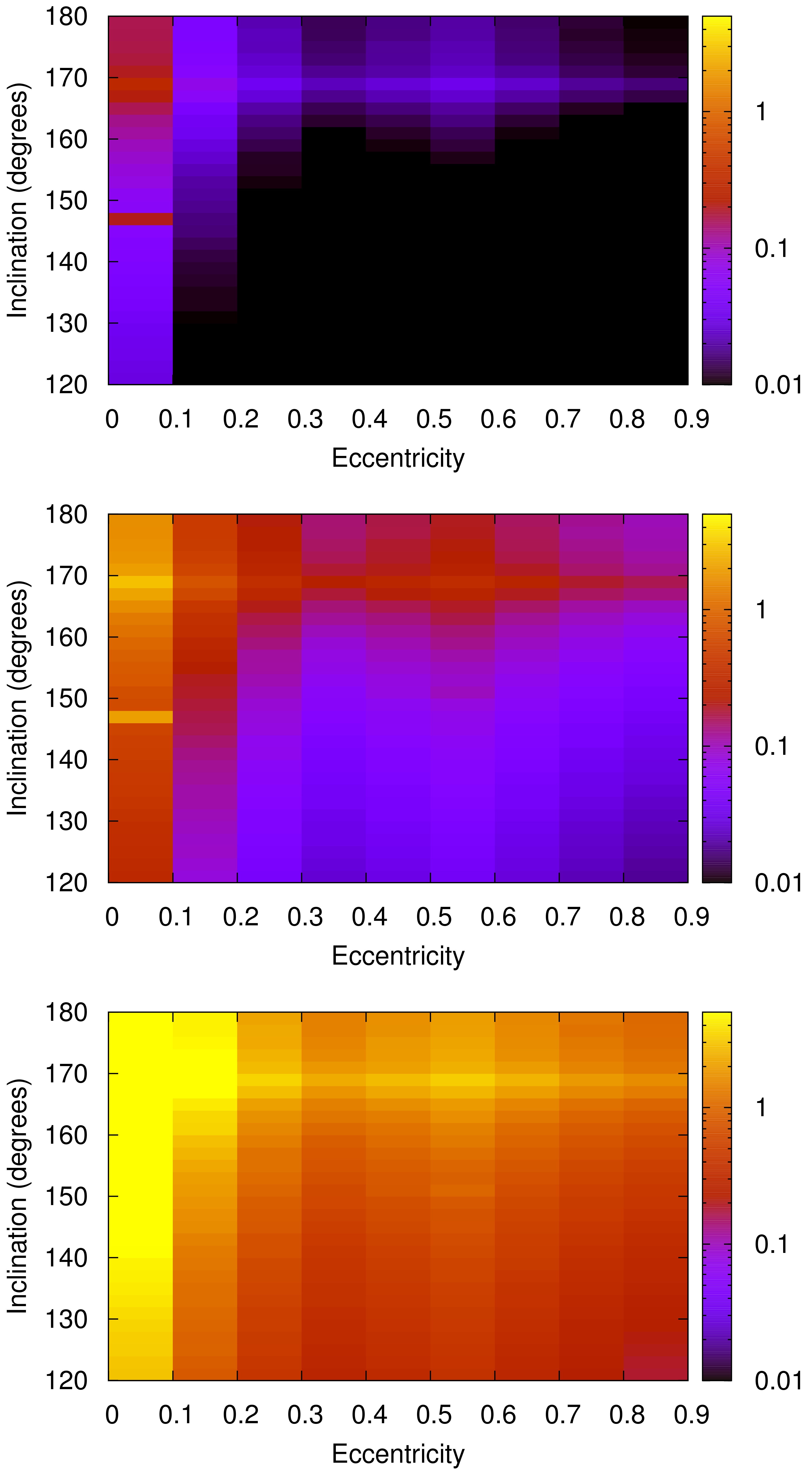}}
\caption{Dust capture efficiency of Iapetus for increasing sizes of the grains (from top to bottom: 0.1, 1 and 10 microns) orbiting Saturn on retrograde orbits. Capture efficiency is expressed as the number of impacts over the migration time across the whole radial path. Dust grains ejected by Phoebe and the other retrograde satellites would reside in the region of the $e-i$ plane ($150^{\circ} < i < 180^{\circ}$ and $e > 0.1$) where Iapetus' capture efficiency is highest.}
\label{cloud-retro}
\end{figure}

We applied our modified model to perfectly absorbing grains of different sizes, which are characterised by different migration timescales through the dependence of $\beta$ by the particle radius (see e.g. p. 35 from \citet{dpl01}). We considered the following values for the size of the grains: 0.1, 1, and 10 $\mu$m. Such range of values was selected on the basis of comparative considerations with the results obtained by \citet{kri02} for the Jovian system. A recent estimate performed by \citet{vsh09} for the Saturnian system suggests more strict constraints (i.e. d $> 3.5$ $\mu$m) to the size of the grains which could survive long enough to be affected by PR drag. The size range we considered overlaps with the one indicated by \citet{vsh09}, therefore we decided to keep our original one since it emphasizes the changes in the capture efficiency as a function of the size of the grains.\\
The estimated sweeping efficiency of Iapetus, expressed through number of collisions during the crossing of the whole radial path, is plotted in Figs. \ref{cloud-pro} and \ref{cloud-retro} for prograde and retrograde dust grains respectively as a function of the eccentricity and the inclination values. While Iapetus has a capture efficiency of $1$ or $100\%$ over those regions of the considered phase space characterized by more than $1$ collision, we preferred to express the capture probabilities as a function of the number of impacts in order to emphasize the different statistical weights of our estimates (i.e. $3$ collisions is statistically more significant than $1$ collision).\\
As can be seen from Fig. \ref{cloud-pro}, the prograde dust grains captured more efficiently are those with very low eccentricity values ($e$ $\approx$ 0): for grains bigger than 10 $\mu$m, the capture probability is of the order of $100\%$ yet, once the dynamical features of the prograde irregular satellites are accounted for, this dynamical class of particles is physically unrealistic. There is a second island of higher efficiency at high eccentricity values ($e$ $>$ 0.4) for those particles having inclination values similar to that of Iapetus but the capture probability is lowest in the region of phase space ($e$ $>$ 0.1 and 30$^{\circ}$ $<$ $i$ $<$ 60$^{\circ}$) populated by the prograde irregular satellites. Set aside for the high-end tail of ejection velocities, the bulk of the impact-generated dust grains would populate the low efficiency region of the $e-i$ plane: as a consequence, the capture efficiency of Iapetus would be of the order of a few 0.1$\%$ for 1 $\mu$m grains and would never exceed 20-30$\%$ even for 10 $\mu$m grains.\\
The situation is different for retrograde grains (see Fig. \ref{cloud-retro}). Again, the highest capture probabilities are in the low eccentricity ($e$ $\approx$ 0), physically unrealistic region. For retrograde grains, however, the island with relatively high capture probability overlap the region of the $e-i$ plane populated by the retrograde irregular satellites (i.e. $e$ $>$ 0.1 and 150$^{\circ}$ $<$ $i$ $<$ 180$^{\circ}$). As a consequence, 1 $\mu$m grains are characterised by capture probabilities ranging between 10-40$\%$, while for the majority of 10 $\mu$m grains the probabilities are of $100\%$.

\subsection{Mass transfer from Hyperion}

Before discussing the conclusions of this work, we would like to review the mass transfer scenarios linking Hyperion to the dark material on Iapetus in light of the results of Cassini mission. Prior to the arrival of Cassini to Saturn, several studies \citep{matt92,mar02,dal04} evaluated the transfer efficiency for collisionally-generated material from Hyperion to Iapetus. \citet{matt92} assumed narrow, conical-shaped clouds of ejecta and estimated a single-passage transfer efficiency of 10$^{-3}$. A more detailed and realistic evaluation of the same scenario performed by \citet{mar02} raised the transfer efficiency to about 20-40$\%$, depending on the characteristics of the break-up event. However, the authors emphasised that the assumptions on the ejection direction were made to study the transfer efficiency of those fragments most likely to reach Iapetus. They also noted that such particles would likely represent only a fraction of the ejecta cloud generated by catastrophic disruption events and that, if isotropic ejection is assumed, the mean transfer efficiency lowers to 0.4$\%$. Such value is in agreement with the one found by \citet{dal04} in studying the impact probability of fragments ejected by Hyperion against the other Saturnian satellites.\\
\citet{matt92} estimated the mass of dark material required to cover the Cassini region on Iapetus to a depth of 1 km to be about 3$\times$10$^{21}$ g (i.e. about half the mass of Hyperion, which is about 5$\times$10$^{21}$ g as reported by \citet{tea07}). The assumption on the depth of the dark material followed from the requirement that no subsequent impact should excavate the whole layer of dark material and expose the bright material underneath. As we said in the introduction, radar measurements of the thickness of the dark material layer performed by Cassini constrained its depth between a few decimetres to about one meter. As a consequence, the amount of material needed to cover the leading hemisphere of Iapetus is at least 3 orders of magnitude smaller than previously thought, i.e. approximately a few 10$^{18}$ g or lower.\\
We can estimate the cratering event needed to roughly supply this amount of material by inverting the formula for the volume of a simple, bowl-shaped crater
\begin{equation}
\label{volume}
 V={\pi}h\left( \frac{d^{2}}{8}+\frac{h^{2}}{6}\right)
\end{equation}
where $h$ is the depth of the crater, $d$ its diameter and $V$ the volume.
If we assume the depth-to-diameter ratio estimated for Hyperion by \citet{tea07}, 0.21$\pm$0.05, we have $d\approx5h$. We then express $V$ as $M\rho^{-1}$ where $\rho$=544 kg m$^{-3}$ (ibid). We assume for $M$ the value computed by \citet{matt92} corrected for the new depth of the dark material on Iapetus: $M$=3$\times$10$^{18}$ g. By inverting eq. \ref{volume} we have
\begin{equation}
 h=\left(\frac{24M}{79\pi\rho}\right)^{\frac{1}{3}}
\end{equation}
This indicates that, if we assume a complete transfer between Hyperion and Iapetus, the needed amount of material would be supplied by excavating a crater with depth $h$=8.11 km and diameter $d$=40.55 km, which is comprised in the crater size distribution observed on Hyperion \citep{tea07}. However, we need to take into account the effects of the transfer efficiency. If we assume the most favourable single-event scenario, i.e. a collimated, high velocity ejection with a 40$\%$ transfer efficiency (the maximum estimated by \citet{mar02}), the resulting crater would have $h$=11 km and $d$=55 km. If we consider a more realistic yet still favourable case with a 20$\%$ transfer efficiency \citep{mar02}, the dimensions of the crater become $h$=13.87 km and $d$=69.33 km, nearing the high-end tail of Hyperion's crater size distribution \citep{tea07}. It should be noted that, in their work, \citet{mar02} assume that the fragments are ejected in the direction that maximises the transfer efficiency; moreover, these author assume a uniform distribution of the ejection velocities with a maximum value of 1.5 km s$^{-1}$, likely overestimating the contribution of high-velocity ejecta. Finally, as emphasised by \citet{tea07}, Hyperion is characterised by a significant porosity, of the order of 42$\%$ if the moon is mainly composed by water ice and likely higher if the rock fraction is significant. For porosity values this high, impacts can be extremely ineffective in excavating high-velocity ejecta since the main crater-forming process is compression instead of excavation \citep{hea02}. As reported by \citet{hea02}, for a target porosity of $60\%$ only $2\%$ of the ejecta achieve velocities greater than 10 m s$^{-1}$. This implies that our estimates represent a lower limit to the size of the crater needed to supply the dark material on Iapetus and that more than one cratering event is likely necessary.\\
Before concluding this review of the scenarios linking Hyperion to the dark material on Iapetus, we would like to point out that the present porosity of Hyperion could be the by-product of a past major collisional event and that the parent body of the satellite could have been characterised by a significantly lower porosity. Under this assumption, if the mass of the parent body was about $10\%$ higher than Hyperion's present mass and such excess mass was collisionally removed, in principle Hyperion could supply the right amount of material observed on Iapetus.


\subsection{Comparative discussion of the mass transfer scenarios}

As we previously said, due to its high porosity the present-day Hyperion is an ineffective source for the dark material coating the leading hemisphere of Iapetus. A most viable scenario connecting Hyperion to the dark material implies a significantly lower porosity of Hyperion's parent body and the collisional removal of about $1/10$ of the present mass of the satellite. However, the excavated material would be delivered in the form of collisional shards and fragments, thus leaving open the issue of how to obtain the uniform dark blanket we now observe on Iapetus.\\
As a comparison, due to its lower porosity and higher density \citep{por05}, Phoebe would be a more efficient source of ejecta. A medium sized crater like Hylas, with its diameter $d$=28 km and its depth $h$=4.67 km computed through its depth to diameter ratio 1:6 \citep{gie06}, would supply 2.43$\times$10$^{18}$ g, i.e. about the estimated amount of dark material. As for Hyperion, not all the material excavated by impacts on Phoebe will be ejected as dust: part of it will consist of collisional shards and fragments. However, the results of \citet{tur08} show that those fragments too big to be influenced by radiation forces would likely re-impact the satellite during their dynamical lifetime. The ejection-reaccretion process would repeat also for the material excavated by these secondary impacts with an ejection velocity higher than Phoebe's escape velocity, i.e. about $100$ $\mathrm {m\,s}^{-1}$ \citep{por05}, with an overall enhanced production of dust.\\
The divergent results we obtained for the visible and infrared data, however, argue against the identification of Phoebe as the sole source of the dark material. Due to the likely active collisional history of the irregular satellites in Saturn system, a multi-source scenario like the one hypothesised by \citet{bea05b} is more plausible and the contribution of different satellites could explain the observed spectral differences between Phoebe and Iapetus. Moreover, if the conjectured nature of Phoebe's gap \citep{tur08} is real, Phoebe collisionally removed once-existing prograde regular satellites near its orbital region and reaccreted most of the generated collisional shards: in such scenario, the contribution to the dust flux of these now-extinct satellites should be accounted for in interpreting the spectral features of the dark material.\\
The existence of a disk around Saturn which spans over the orbital region of the irregular satellites argues in favour of this scenario. First, it proves that collisional, dust-generating processes acted and are likely still acting in the outer Saturnian system. Second, the lower limit to its mass estimated by \citet{vsh09} under the conservative assumption that the disk is composed by $10$ $\mu$m grains indicates that even moderate cratering events (i.e. producing craters of about $1$ km in diameter) would be able to supply enough material to form such disk. Assuming the present disk is in a steady state, the accumulation rate on Iapetus would be of $20$ $\mu$m Myr$^{-1}$ \citep{vsh09}. Depending on the cratering rate on Iapetus (see e.g. the timescale estimated by \citet{zea03} for ecliptic comets) and the real size distribution of the grains composing the disk, such rate can possibly be enough to resupply the material excavated on Iapetus by impacts.\\

\section{Conclusions}

The spectral range covered by Cassini/VIMS and the availability of data from the close flyby of Iapetus allowed us to perform an automatic spectral classification of the surfaces of Iapetus, Hyperion and Phoebe at comparably high spatial resolution and favourable geometry. Our aim was to look for spectral affinities between the dark hemisphere of Iapetus and the two satellites that have been historically indicated as possible sources of the exogenous material covering it. The classification has been performed separately and adopting different approaches for the visual and infrared portions of the spectra measured by VIMS. For the former, the I/Fs measured in all the available wavelengths were used as variables, while several diagnostic features and the spectral continuum were used as variables for the latter, further separating the results obtained with the application of two different sensitivity functions in the calibration pipeline.\\
As a general remark, striking spectral associations between Iapetus' dark material and Phoebe or Hyperion are hardly found in these data. In the visible range, the G-mode analysis confirms that Phoebe is essentially grey, with a subset of the spectra even showing a negative slope, while Iapetus and Hyperion are clearly reddened. However, we found no association between the dark side of Iapetus and Hyperion, since the reddening appears stronger on Hyperion than on Iapetus while the dark material on Iapetus shows a lower albedo than Hyperion. Moreover, any association in the visible range would not be conclusive, as similar photometric correlations have been pointed out also with other small dark satellites moving on both retrograde and prograde orbits \citep{bea05b}. In the near infrared range the correlation appears generally stronger between Iapetus and Phoebe, although significant variability of some compounds is found on all the three satellites. We observed a clear correlation of most non-ice features with water ice in the case of Hyperion, which suggests that these compounds exist as trapping structures (e.g., fluid inclusions) into water ice, whereas the same kind of correlation - particularly for spectral features clearly correlated with dark material - is not present in the data on Iapetus and Phoebe. Furthermore, the spectra of Hyperion in the near infrared show a continuum profile significantly different from those of Iapetus and Phoebe.\\
The evaluation of Iapetus' sweeping efficiency in a PR drag based scenario yielded interesting results. The unusually high orbital inclination of Iapetus naturally enhances its capture efficiency for retrograde dust grains and lower that for prograde grains. Statistically, a significant fraction of the retrograde dust particles impacts the satellite while migrating inward for grain sizes greater than 1 $\mu$m. For grains of several microns in size, i.e. those that would not be removed by radiative forces as reported by \citet{vsh09}, this fraction approaches unity. For prograde particles, on the contrary, the transfer efficiency is an order of magnitude lower and even for the biggest grains we considered it never exceeds $10-30\%$. Together with the hypothesised past history of the Saturnian irregular satellites, i.e. the formation of Phoebe's gap \citep{tur08,tur09}, the capture of Phoebe \citep{tur09} and the creation of collisional families \citep{nes03,tur09}, this result can explain the striking appearance of the satellite. The interpretation of ultraviolet data suggests that the delivery of the dark material is a recent or even still ongoing process \citep{hah08}. This can be naturally explained in the PR drag scenario through the effects of the impacts of micrometeorites on the irregular satellites. Such micro-impacts, particularly those on the smaller retrograde satellites characterised by lower escape velocities, would supply a continuous source of dust as suggested by the data of Galileo mission on the Jovian system \citep{kri02}. The existence of the newly discovered outer disk around Saturn \citep{vsh09} strongly supports this scenario.\\
In conclusion, the results of our work argue in favour of a link between Phoebe and the dark material on Iapetus due to Poynting-Robertson drag of dust particles both on spectroscopic and dynamical basis. While likely being the main actor, we argue that Phoebe was not the sole source which contributed in supplying the dark material. Collisionally removed prograde satellites which originally populated the orbital region near Phoebe and the parent bodies of the suggested collisional families existing in the system likely played a role and could account for the differences observed in the spectroscopic data.


\bsp

\label{lastpage}

\end{document}